\documentstyle[floats,preprint,aps,epsf]{revtex}
\tightenlines

\newcommand{\hs}[1]{\hspace*{#1}}

\newcommand{\cov}{\bigtriangledown}

\newcommand{\beq}{\begin{equation}}
\newcommand{\eeq}{\end{equation}}
\newcommand{\beqa}{\begin{eqnarray}}
\newcommand{\eeqa}{\end{eqnarray}}
\newcommand{\sr}{\sqrt}
\newcommand{\fr}{\frac}
\newcommand{\mn}{\mu \nu}

\def\vereq#1#2{\lower3pt\vbox{\baselineskip1.5pt \lineskip1.5pt
\ialign{$\m@th#1\hfill##\hfil$\crcr#2\crcr\sim\crcr}}}

\begin{document}

\draft
\preprint{ KEK-TH-803, hep-th/0202147}

\title{On the Stability of Black Strings/Branes\footnote{Based on
a talk given at the 11th Workshop on General Relativity and Gravitation 
held at Waseda University in Japan during Jan. 9-12, 2002}}
\author{ Gungwon Kang\footnote{E-mail address:
gwkang@post.kek.jp}} 
\address{ Theory Group, KEK, Tsukuba, Ibaraki 305-0801, Japan}
\maketitle
\begin{abstract}

Some issues on the stability of black string or brane solutions 
are summarized briefly. The stability of dS/AdS-Schwarzschild 
black strings has been investigated. Interestingly, the 
AdS-Schwarzschild black strings turn out to be stable as the 
horizon size becomes larger than the AdS scale. It is also 
shown that BTZ black strings in four dimensions are stable 
regardless of the horizon size. Such stable feature seems to be 
common for several known black strings in dimensions lower than 
five. Some implications of our results on the role of 
non-uniformity in stable black string configurations are also 
discussed.

\end{abstract}



\section{Introduction}

Black holes are solutions of Einstein gravity, which possess 
usually a compact null hypersurface with topology of sphere. 
Black strings or branes are a sort of higher dimensional 
generalizations of black holes. The simplest black strings or 
branes are the product of Schwarzschild black holes and an 
infinite line or a $p$-dimensional plane. The extra dimensions 
can also be compactified. Black string/brane solutions arise
naturally in string theory as well. 
One of the most important reasons that one considers black hole 
solutions seriously, inspite of their odd causal structure, 
is that they are stable. The stability of black strings/branes
was studied by Gregory and Laflamme~\cite{GL}. They showed that
black strings/branes are generically unstable to linearized 
perturbations with a large wavelength along the string.
For example, the four-dimensional Schwarzschild black hole cross 
a circle of length $L$ becomes unstable under linearized 
perturbations as $L$ becomes larger than the order of the 
Schwarzschild radius $r_{\rm h}$. 

One naive explanation for this instability is the entropy 
comparison with that of a black hole with the same mass. Namely,
a black hole configuration is entropically preferable as the size 
of the black string $L$ increases. However, the idea of assigning 
entropy to black holes or strings is based on the quantum behavior 
of such black objects which is not neccessarily related to the 
classical stability behavior of the black string. Moreover, this 
argument of entropy comparison is global in nature. 
Recently, Gubser and Mitra have refined this global thermodynamic
analysis by considering the local thermodynamic properties of the
black object~\cite{GM}. They conjectured that a black string/brane
with noncompact translational symmetry is classically stable if,
and only if, it is locally thermodynamically stable. A slightly
modified version of this conjecture applicable even for compact 
cases is recently stated by Hubeny and Rangamani~\cite{HR}.
The proof of the Gubser-Mitra (GM) conjecture is sketched by 
Reall~\cite{Reall}, and a more complete illustration in the simplest
case of Schwarzschild black strings/branes is given by Gregory and 
Ross~\cite{GR}. The GM conjecture practically provides a very
powerful and easy way to check whether given black string/brane 
solutions are stable or not, compared to the usually complicated 
numerical analysis of classical linearized perturbations. 

The black string instability found at the linear level was believed
to indicate that the full nonlinear evolution of the instability
would result in the fragmentation of the black string. In order to 
be consistent with the classical theorem of no bifurcation of the 
event horizon, presumably quantum effects will play some important 
role when the shrinkage of the string horizon reached to the stage 
of large enough curvature. This widely accepted idea of black 
string fragmentations has been used in many discussions in the 
literature, involving black string/brane configurations. Recently, 
however, Horowitz and Maeda have shown that black strings do not 
break up within a finite affine time in their full classical 
nonlinear evolution~\cite{HM}. Thus, unstable black strings are 
likely to evolve into black string-like configurations which are 
probably non-uniform and stable even under linearized 
perturbations~\cite{HM2}. In addition, the recent numerical study 
about approximate non-uniform black string solutions by 
Gubser~\cite{Gubser} indicates that such transition is not 
continuous, but the first order in thermodynamic nature. 

In order to understand the black string instability better, it will 
be of interest to see whether or not stable black string/brane 
solutions exist, and, if it does so, to understand what makes them 
stable. The only known black string/brane solutions even when 
$L\geq r_h$ are the extremal black $p$-branes carrying certain 
charges~\cite{GLcharge}. 
In this talk, I present some black string/brane solutions that 
reveal stable behavior under linearized perturbations. One type of 
them is AdS-Schwarzschild black strings in anti-de Sitter 
space with/without a uniform tension brane~\cite{HK}. These black 
strings/branes become stable as the horizon radius becomes larger 
than the order of the AdS radius of the background geometry 
perpendicular to black strings/branes. The other type is black 
string/brane solutions in the spacetime dimensions lower than 
five~\cite{HKL}. In particular, the BTZ black strings in four 
dimensions turn out to be stable {\it always}, regardless 
of the transverse horizon size. Some physical implications of our 
results and some open issues are also discussed.

\section{Stable Black Strings in Anti-de Sitter Space}
\label{AdS}

In this section, I briefly summarise the results in Ref.~\cite{HK}
and give some further results obtained. In the five-dimensional 
Einstein gravity with negative cosmological constant 
$\Lambda_5 =-6/l^2_5$, one can have black string solutions whose
metrics are given by
\beqa
ds^2 &=& H^{-2}(z) (\gamma_{\mn}dx^{\mu}dx^{\nu} + dz^2 ), 
\label{metric0}    \\
&=& H^{-2}(z) \Big[ -f(r)dt^2 +\frac{1}{f(r)}dr^2 
+r^2d\Omega_2^2 +dz^2 \Big] ,
\label{bsmetric}
\eeqa
where the warping factors are
\beq
H(z) = \left\{\begin{array}{ll}
              l_4/l_5 \sinh z/l_4, & dS_4 (\Lambda_4 > 0) \\
              z/l_5, & M_4 (\Lambda_4=0) \\
              l_4/l_5 \sin z/l_4, & AdS_4 (\Lambda_4 < 0), 
              \end{array}
         \right.
\label{warps}
\eeq
and $f(r)= 1-r_0/r-\Lambda_4r^2/3$. 
Here the four-dimensional cosmological constant $\Lambda_4 = 
\pm 3/l^2_4$ is arbitrary. When $r_0 =0$, these metrics actually 
describe the same five-dimensional pure anti-de Sitter spacetime 
(${\rm AdS}_5$), and simply correspond to different ways of slicing it, 
e.g., de Sitter, flat, and anti-de Sitter slicings. If a 3-brane 
with uniform tension is introduced at $z=0$, the resulting 
geometries are still described by the metrics above with replacing 
$z \rightarrow |z| +c$ and now $\Lambda_4$ is determined by the 
location and the tension of the 3-brane accordingly. The stability
for the flat case (i.e., $\Lambda_4=0$) has been already studied 
in Refs.~\cite{Gcosmo,CHR}. 

In order to check the linearized stability, let us consider small metric 
perturbations, $g_{MN}(x) \rightarrow g_{MN}(x)+h_{MN}(x,z)$, about 
these black string background spacetimes, and see whether or not there 
exists any mode which is regular spatially, but grows exponentially in 
time. As in the usual Kaluza-Klein reduction, the massive scalar 
($h_{zz}$) and vector ($h_{\mu z}$) fluctuations can always be set to 
be zero by using the five-dimensional diffeormorphism symmetry, 
leaving only four-dimensional {\it massive} spin-2 gravity fluctuations. 
Moreover, the scalar and vector components of the five-dimensional 
linearized perturbation equations reduce to the four-dimensional 
transverse traceless gauge conditions for the massive spin-2 
fluctuations $h_{\mu\nu}$~\cite{Kang,Reall}. By putting $h_{\mu\nu}(x,z) 
= H^{3/2}(z)\xi (z) h_{\mu\nu}(x)$, the linearized equations turn out
to be
\begin{eqnarray}
\mbox{} & & \Delta_L h_{\mu\nu}(x)
\equiv\Box h_{\mu\nu}(x) + 2R_{\mu\rho\nu\tau}h^{\rho\tau}(x)
  =m^2 h_{\mu\nu}(x),  
\label{lich}  \\
& & \cov^{\mu}h_{\mu\nu} = 0, \hs{3ex} h=\gamma^{\mu\nu}h_{\mu\nu}=0,
\label{gc}  \\
& & \big[ -\partial_z^2 +V(z) \big]\xi (z) = m^2\xi (z), \hs{3ex}
V(z)=-\frac{3}{2}\fr{H''}{H}+\fr{15}{4}\Big(\fr{H'}{H}\Big)^2 .
\label{KKmass}
\end{eqnarray}

Notice first that, even if the black strings we are considering are
not uniform due to the warping factors along the string, the 
perturbation equations above become separable as in the case of 
translationally invariant black strings. The only difference is the
spectrum of the Kaluza-Klein (KK) mass with the non-vanishing 
effective potential $V(z)$ in Eq.~(\ref{KKmass}).
When $m^2=0$, as is pointed out in Ref.~\cite{GL}, 
Eqs.~(\ref{lich}) and (\ref{gc}) are exactly same as those for
perturbations about four-dimensional black hole spacetimes, which are 
known to be stable. Since adding a mass term 
usually increases stability, it had been believed for some time that
black strings are stable. However, it turns out that the extra degrees
of freedom coming from the massiveness could give unstable 
solutions~\cite{GL,Kang}.

\begin{figure}[tbp]
 \centerline{\epsfxsize=85mm\epsffile{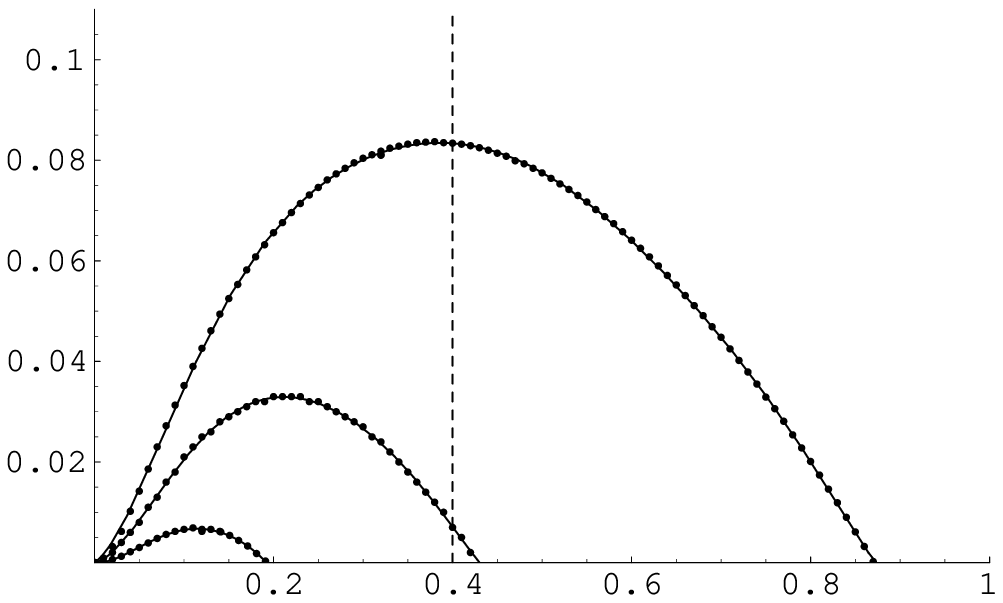}
  \epsfxsize=85mm\epsffile{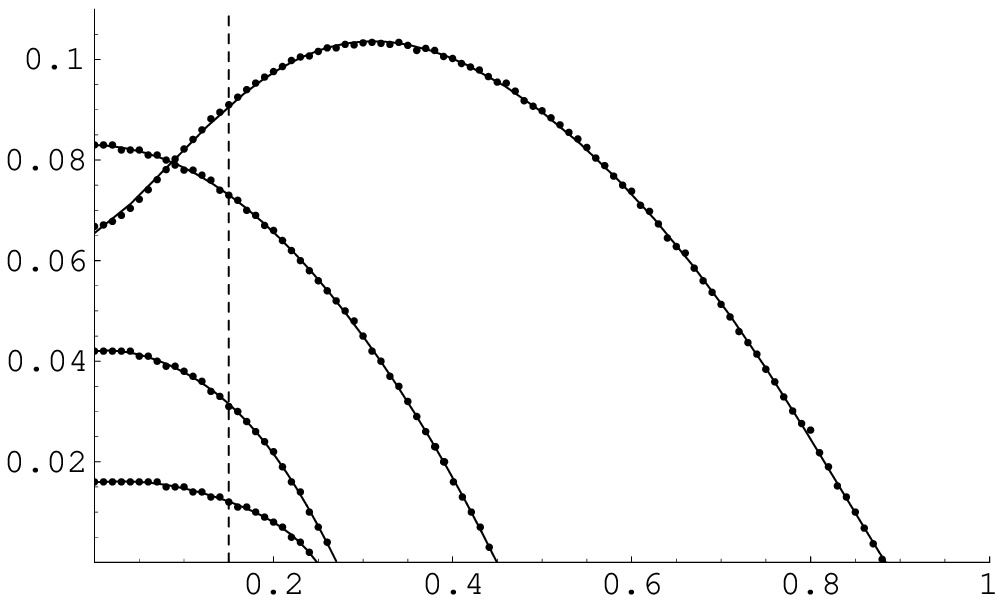}
   \put(-215,135){$\Omega$}
   \put(-130,146){$dS$}
   \put(-15,15){$m$}
   \put(-70,80){$r_0=1$}   
   \put(-140,60){$r_0=2$}   
   \put(-210,64){$r_0=3.5$}   
   \put(-218,34){$r_0=3.8$}   
   \put(-400,146){$AdS$}
   \put(-460,135){$\Omega$}
   \put(-320,78){$r_0=1$}   
   \put(-420,56){$r_0=2$}   
   \put(-440,26){$r_0=4$}   
   \put(-260,15){$m$}
 }
 \caption{The left figure is for the $AdS$ case with $r_0=1, 2$ and $4$.
 The right figure is for the dS case with $r_0=1, 2, 3.5$, and $3.8$. 
 The Nariai solution corresponds to $r_0 \simeq 3.85$. The fixed AdS 
 and dS radius is $l_4=10$. The straight vertical lines denote the 
 lowest KK masses, $0.4$ for AdS and $0.15$ for dS.}
 \label{fig1}
\end{figure}

The strongest instability is expected for the $s$-wave fluctuations.
General, spherically symmetric perturbations can be written in 
canonical form as~\cite{VRW,GL}
\begin{eqnarray}
h_{\mu\nu} (x) &=& e^{\Omega t}
  \left(\begin{array}{cccc}
   H_{tt}(r) &H_{tr}(r) &0 &0 \\
	 H_{tr}(r)&H_{rr}(r)&0&0\\
	 0&0&K(r)&0\\
	 0&0&0&K(r) \sin^2\theta
	\end{array}\right) ~,
\end{eqnarray}
with $\Omega >0$.
From the coupled equations in Eqs.~(\ref{lich}) and (\ref{gc}), 
we can eliminate all but one variable, say $H_{tr}$, obtaining 
a second order ordinary differential equation in the following form:
\beq
A(r; r_0, \Lambda_4, \Omega^2, m^2)H''_{tr} +BH'_{tr} +CH_{tr} =0 .
\label{mastereq}
\eeq
With suitable boundary conditions (see the details in Ref.~\cite{HK}),
one can solve this equation numerically, and unstable solutions for 
the {\it massive} spin-2 fluctuations in Eq.~(\ref{lich}) are shown in 
Fig.~\ref{fig1} for given dS/AdS-Schwarzschild black hole 
background spacetimes. 

As shown in Ref.~\cite{HK} explicitly, the most important observation 
is that adding a negative cosmological constant $\Lambda_4$ has a 
stabilization effect. As the horizon size increases, the instability 
becomes weak as in the case of Schwarzschild black holes~\cite{Gcosmo}. 
For the AdS case, however, the threshold mass 
$m_* \equiv m(r_0, \Lambda_4, \Omega =0)$ vanishes even at a finite 
$r_0$, not as $r_0 \rightarrow \infty$. The numerical search shows
that this termination of unstable solutions occurs approximately at
$r_0 \simeq 0.77l_4$ (i.e., $r_+ \simeq 0.58l_4$). In addition to 
it, the KK mass spectrum determined by Eq.~(\ref{KKmass}) has a 
finite mass gap as indicated by the vertical line in 
Fig.~\ref{fig1}. Therefore, it turns out that 
${\rm AdS}_4$-Schwarzschild black strings in ${\rm AdS}_5$ space 
become stable when the horizon size is $r_+ > r^{\rm cr}_{+} 
\simeq 0.20 l_4$. The presence of a 3-brane as in the brane world 
model~\cite{RS} simply increases this critical value. 
For the dS case, on the other hand, the threshold masses remain 
larger than the lowest KK mass for all horizon radii bounded by the 
cosmological horizon. In addition, since the KK mass spectrum is 
continuous, all ${\rm dS}_4$-Schwarzschild black strings in 
${\rm AdS}_5$ space are unstable. 

The local thermodynamic stability of a segment of AdS/dS black 
string will be determined by the sign of the heat capacity given by 
$dM/dT \sim -2\pi r^2_+ (1-\Lambda_4r^2_+)/(1+\Lambda_4r^2_+)$.
For the AdS case, one can easily see that the heat capacity becomes 
positive for $r_+ > l_4/\sr{3}$. Thus, we expect AdS black strings
become stable classically when $r_+ > l_4/\sr{3} \simeq 0.58 l_4$ 
according to the GM conjecture. The slight difference in the critical
values is expected because of the non-uniformity of the AdS black 
string due to the warping factor~\cite{HK}.
On the other hand, dS black strings are expected to be
unstable classically since the heat capacity is always negative for 
all $r_+ < l_4/\sr{3}$ within the cosmological horizon.

\section{Lower Dimensional Black Strings/Branes}
\label{LD}

It was argued in Ref.~\cite{EHM} that the instability  of BTZ black 
strings in the four-dimensional brane worlds sets in when the 
transverse size of the black string reaches the AdS scale. Here we 
show that this naive expectation, which is based on the entropy 
comparison with a localized black hole around a 2-brane with same 
mass, is not true. Surprisingly, they turn out to be stable always. 
The metric of rotating BTZ black strings in four dimensions can be 
written by
\beq
ds^2 = H^{-2}(z) \left[ -fdt^2 +\fr{dr^2}{f} +r^2(d\varphi 
-\fr{J}{2r^2}dt)^2 +dz^2 \right] ,
\eeq
where $f(r)=-M +r^2/l^2_3 +J^2/4r^2$ and $H(z)=l_3/l_4 \sin z/l_3$.
The first evidence for the stable behavior can easily be obtained
by applying the GM conjecture. The local thermodynamic stability of 
a segment of this black string is governed by that of the 
three-dimensional BTZ black hole. The heat capacities are given by 
\beqa
C_J & \equiv & T(\partial S/\partial T)_J
= 4\pi r_+ (r^2_+ -r^2_-)/(r^2_+ +3r^2_-) , \\
C_M & \equiv & T(\partial S/\partial T)_M 
= 4\pi r_+ (r^2_+ -r^2_-)/(3r^2_+ +r^2_-), 
\eeqa
both of which are always positive since the inner horizon radius is 
$r_- \leq r_+$. If a thermal system has rotations, the thermodynamic 
stability requires $(\partial \Omega_H/\partial J)_T \geq 0$ 
additionally since adding angular momentum is expected to increase 
the angular velocity of the system in average. For the rotating BTZ 
black hole, it turns out that 
\beq
\left( \fr{\partial \Omega_H}{\partial J}\right)_T 
= \fr{\Omega_H}{J} \fr{r^2_+ -r^2_-}{r^2_+ +3r^2_-}
\label{rotchemi}
\eeq
which is also positive always.
According to the GM conjecture, therefore, we expect that the rotating
BTZ black string in four dimensions is stable under linearized 
perturbations, regardless of the horizon size. 

Now for the linearized perturbation analysis, the equations can be 
separated as before, and the most general form of the three-dimensional 
metric fluctuations is given by~\cite{HKL}
\begin{eqnarray}
h_{\mu\nu} (x) &=& e^{\Omega t +i{\rm n}\varphi}
  \left(\begin{array}{ccc}
   H_{tt}(r) &H_{tr}(r) &i{\rm n}H_{t\varphi}(r)  \\
	 H_{tr}(r)&H_{rr}(r)&0  \\
	 i{\rm n}H_{t\varphi}(r) &0 &r^2K(r) \\
	\end{array}\right) ~.
\end{eqnarray}
The decoupled equation is in the same form as Eq.~(\ref{mastereq}),
but it becomes much complicated and a complex equation due to the
rotation parameter $J \not = 0$. For static BTZ black holes (i.e., 
$J=0$), however, one can set $H_{t\varphi}=0$ by using the gauge 
freedom and easily solve the equation numerically. We could not 
find any unstable mode for various values of $M$ and $l_3$~\cite{HKL}.
That is, {\it massive} spin-2 fluctuations do not seem to possess any
instability for the {\it three-dimensional} BTZ black hole 
background. Therefore, we expect that the static BTZ black strings
in four-dimensions are always stable under linearized perturbations.
This result agrees well with the local thermodynamic stability 
through the GM conjecture.

\section{Discussion}

We have shown briefly that the AdS-Schwarzschild black strings in
five dimensions become stable under linearized perturbations as the
horizon size parameter (i.e., $r_+$) becomes larger than the order of 
the ${\rm AdS}_4$ radius $l_4$. The higher dimensional extension of 
this result is straightforward. The AdS black strings in 
$(4+n)$-dimensions can be obtained simply by replacing $f(r) 
\rightarrow 1 -(r_0/r)^n +r^2/l^2_{3+n}$ and $d\Omega^2_2 
\rightarrow d\Omega^2_{1+n}$ in Eq.~(\ref{bsmetric}). One can easily 
see that the sign of the heat capacity for such systems also becomes 
positive as the horizon size increases. For black $p$-branes 
(i.e., $dz^2 \rightarrow \delta_{ij}dz^idz^j$ for $i,j =1, 
\cdots ,p$), the basic feature of the stability is also same.

It should be pointed out that the essential reason for having stable
AdS black strings is that ${\rm AdS}_4$-Schwarzschild black hole 
bacgrounds do not allow unstable massive spin-2 fluctuations as the 
horizon size increases. Furthermore, a sort of effective 
compactification due to the warping geometry along the string
increases its stability in addition. Namely, the warping factor for 
the AdS case rises up as one approaches to the ${\rm AdS}_5$ horizon
whereas those for flat and dS cases go down to zero. Consequently, 
the potential in Eq.~(\ref{KKmass}) for the AdS case becomes box-like 
with a finite mass gap, giving a confining effect for waves along the 
string. Notice that the proper lengths of black strings in the extra 
direction are infinite for all three cases. It is interesting to see 
that the scale of this effective compactification of AdS black 
strings is governed by the ${\rm AdS}_4$ 
scale $l_4$, instead of $l_5$, because the ${\rm AdS}_5$ scale does 
not enter in Eq.~(\ref{KKmass}) at all. However, note that, when
the instability sets in, the {\it minimum} horizon size of the 
AdS black string in {\it proper length} is indeed of order the 
${\rm AdS}_5$ scale, i.e., $r_+^{\rm PL} = r_+^{\rm cr} 
H^{-1}_{\rm min.} = r_+^{\rm cr}l_5/l_4 \simeq 0.20 l_5$, which is 
consistent with the instability mechanism argued in Ref.~\cite{CHR}. 

As mentioned above, it is suggested by Horowitz and Maeda~\cite{HM} 
that, even in the case that a black string is unstable under 
linearized perturbations, its full non-linear evolution presumably 
ends up with a sort of non-uniform black string configuration, 
instead of fragmentations of the string horizon. This final state
must be stationary and stable. Then it will be interesting to see 
what makes this configuration stable even under linearized 
perturbations. In other words, what is the role of the 
``non-uniformity''? The perturbation equations will in general 
consist of two parts; that is, fluctuations for black holes on 
``transverse'' slicings of the black string and that along the 
string. Of course, they become separable perhaps only for special
cases such as uniform black strings and those ones with overal 
warping factors. Now our study above seems to indicate that the 
stability of spin-2 fields with {\it ``bigger'' degrees of freedom} 
on black hole backgrounds of slicings is crucial generically for 
the stability of a non-uniform black string. However, even if such 
fields have ``instability'', some effective compactification due 
to the curvature or periodicity along the black string could also 
make the black string stable. 

For the BTZ black strings in four dimensions, we have shown that 
they are stable {\it independent of the transverse horizon size} 
since the massive spin-2 fluctuation equation itself does not 
possess unstable modes at all. It seems that this stable behavior 
for black strings/branes in dimensions lower than five is common. 
The thermodynamic analysis for several such 
solutions~\cite{HH,Lemos,NK} shows that all of them are locally 
thermodynamically stable~\cite{HKL}. Consequently, the GM conjecture 
implies that they are classically stable under linearized 
perturbations. This special property might be related to the fact 
that gravitational waves in lower dimensions than four do not have 
enough degrees of freedom for propagations. However, we point out 
that the {\it massive} spin-2 fluctuations in lower dimensions must 
have the same number of degrees of freedom as that of the 
four-dimensional massless gravitational waves. 
Further study is required concerning this issue.  

Since the discovery of the Gregory-Laflamme instability in 1993, 
it is somewhat surprising that no specific work has been done for 
the stability of Kerr black strings. This is probably because 
the wide belief was that such rotating black string is also
unstable. However, a naive application of the GM conjecture shows 
that this may not be the case. For instance, the heat capacity of 
the Kerr black hole is negative, but becomes positive as the 
angular momentum parameter increases towards the extremal value. 
Thus, the Kerr black string with large angular momentum parameter 
might be stable under linearized perturbations possibly due to the 
``centrifugal'' effect of rotations in the background geometry.
On the other hand, however, another thermodynamic quantity 
(e.g., $\left( \partial \Omega_H/\partial J \right)_T$ as in 
Eq.~(\ref{rotchemi})) becomes negative exactly for the values of 
angular momentum where the heat capacity is positive. 
The direct analysis of linearized fluctuations in 
Eqs.~(\ref{lich})-(\ref{KKmass}) with $H(z)=1$ is not available at 
the present since the equations are not easily decoupled in this 
Kerr background geometry. We expect that a partial application of 
the four-dimensional Newman-Penrose formalism to this 
five-dimensional system may decouple the set of equations, 
resulting in a sort of ``Teukolsky equations''~\cite{KangKerr}. 
The detailed analysis of this decoupled equation will reveal all 
non-trivial behaviors for the stability of Kerr balck strings in 
the parameter space of mass and angular momentum. Consequently, 
it will also hint at how to extend Reall's proof of the GM 
conjecture to a system with {\it rotation}. 

Finally, it is also very interesting to extend the proof in 
Ref.~\cite{HM} to the cases of Schwarzschild or dS-Schwarzschild
black strings in ${\rm AdS}_5$. It is because such spacetimes 
have naked singularities at the ${\rm AdS}_5$ horizon outside
the event horizon.

\section*{Acknowledgments}

The author would like to thank T. Hirayama, K. Lee, Y. Lee, R. Myers, 
and P. Yi for useful discussions. This work was supported by JSPS 
(Japanese Society for Promotion of Sciences) Postdoctoral 
Fellowships.

\end{document}